\documentclass[acmtog]{acmart}
\acmSubmissionID{211}

\usepackage{booktabs} % For formal tables
\usepackage{amsmath}
\usepackage{amsfonts}
\usepackage{amssymb}
\usepackage{mathtools}
\usepackage{float}
\usepackage[linesnumbered,ruled,vlined]{algorithm2e}
\usepackage[toc,page]{appendix}

\usepackage[ruled]{algorithm2e} % For algorithms

\SetAlFnt{\small}
\SetAlCapFnt{\small}
\SetAlCapNameFnt{\small}
\SetAlCapHSkip{0pt}

\newcommand{\ours}{ManifoldPlus}

\begin{document}
% Title portion
\title{\ours{}: A Robust and Scalable Watertight Manifold Surface Generation Method for Triangle Soups}

\author{Jingwei Huang}
\affiliation{
  \institution{Stanford University}
}
\author{Yichao Zhou}
\affiliation{
	\institution{University of California, Berkeley}
}
\author{Leonidas Guibas}
\affiliation{
	\institution{Stanford University}
}


\begin{abstract}
We present \ours{}, a method for robust and scalable conversion of triangle soups to watertight manifolds.
While many algorithms in computer graphics require the input mesh to be a watertight manifold, in practice many meshes designed by artists are often for visualization purposes, and thus have non-manifold structures such as incorrect connectivity, ambiguous face orientation, double surfaces, open boundaries, self-intersections, etc.
Existing methods suffer from problems in the inputs with face orientation and zero-volume structures. Additionally most methods do not scale to meshes of high complexity.
In this paper, we propose a method that extracts exterior faces between occupied voxels and empty voxels, and uses a projection-based optimization method to accurately recover a watertight manifold that resembles the reference mesh.
Compared to previous methods, our methodology is simpler. It does not rely on face normals of the input triangle soups and can accurately recover zero-volume structures.
Our algorithm is scalable, because it employs an adaptive Gauss-Seidel method for shape optimization, in which each step is an easy-to-solve convex problem.
We test \ours{} on ModelNet10~\cite{wu20153d} and AccuCity\footnote{\url{https://www.accucities.com}} datasets to verify that our methods can generate watertight meshes ranging from object-level shapes to city-level models.
Furthermore, through our experimental evaluations, we show that our method is more robust, efficient and accurate than the state-of-the-art. Our implementation is publicly available\footnote{\url{https://github.com/hjwdzh/ManifoldPlus}}.
\end{abstract}

\begin{teaserfigure}
    \centering
    \includegraphics[width=\textwidth]{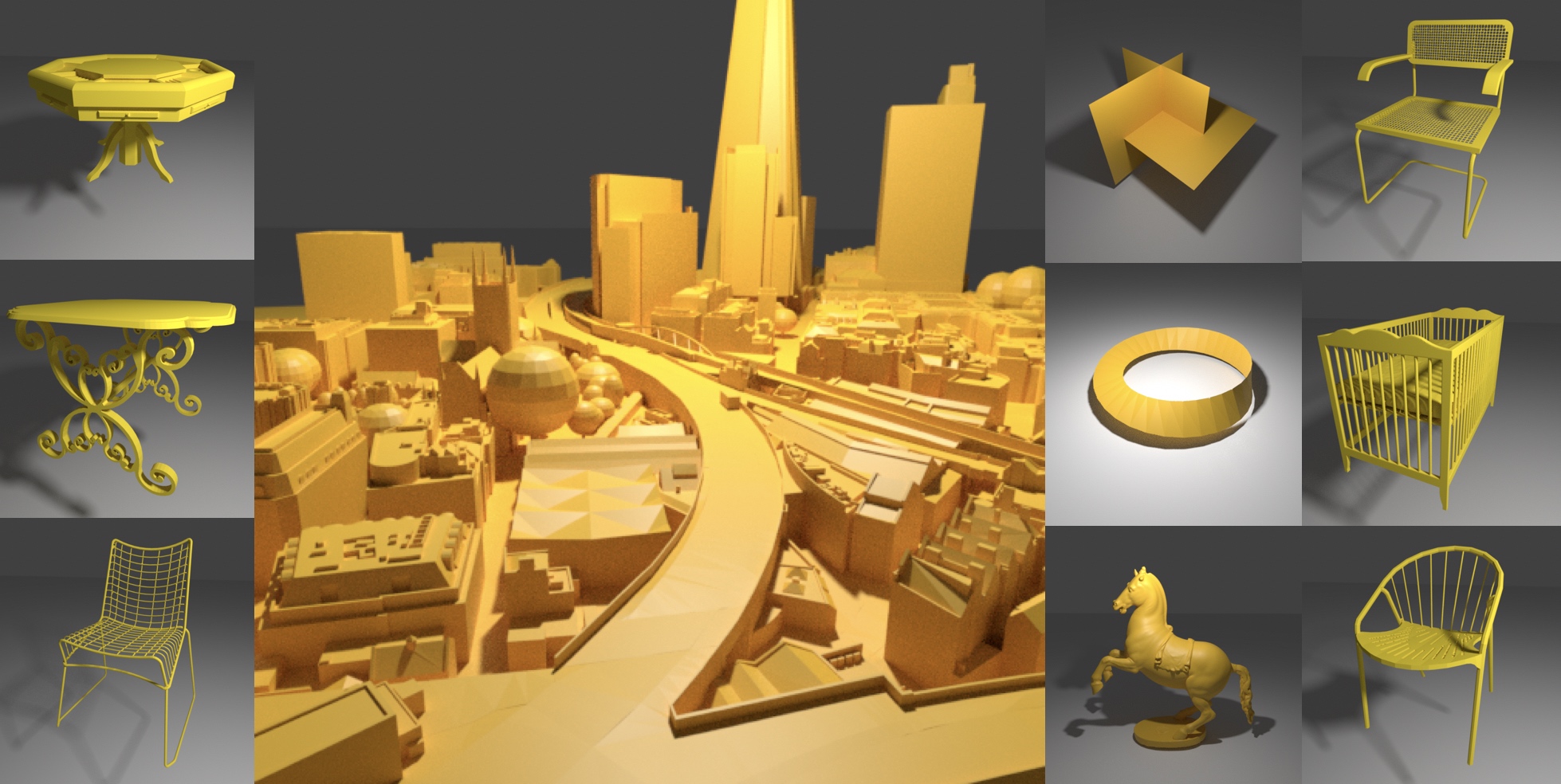}
    \caption{\ours{}: We present a method to robustly convert complex meshes in larget scale to watertight manifolds.}
    \label{fig:teaser}
\end{teaserfigure}

\maketitle

\section{Introduction}
\begin{figure}
\includegraphics[width=\linewidth]{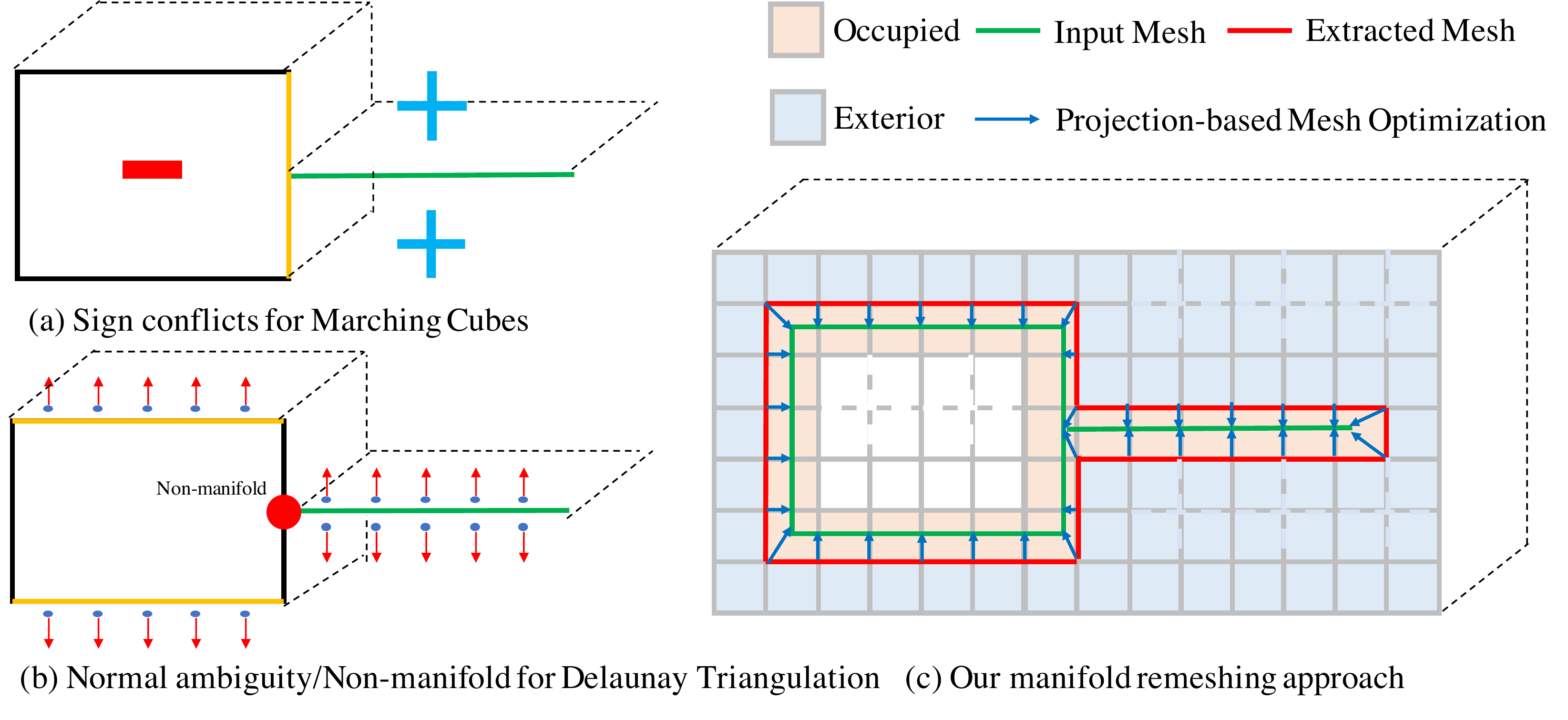}
\caption{Illustration of manifold remeshing challenge. (a) Marching cubes cannot resolve sign conflicts at T-junctions. (b) There is natural ambiguity regarding exterior information. Non-manifold edges can be created at T-junctions by removing all exterior regions. (c) Our novel method addresses these problems.}
\label{fig:intro-thin}
\end{figure}

% introduce the problem
Watertight manifolds are ubiquitous in computer graphics. They are usually represented as \textit{orientable 2-manifold} triangle meshes. Such topology gives a convenient surface representation of objects where local geodesic neighborhoods can be consistently analyzed.  Various tasks in computer graphics often mandate or at least prefer a watertight manifold mesh as input, including geometry-based segmentation~\cite{kim1992recognition,gelfand2004shape}, quadrangulation~\cite{jakob2015instant,huang2018quadriflow}, UV mapping~\cite{burley2008ptex,poranne2017autocuts}, mesh deformation~\cite{sorkine2007rigid,uy2020deformation}, physically-based simulation~\cite{baraff1998large,wu2001adaptive}, finite element analysis~\cite{zienkiewicz1977finite}, surface feature extraction for machine learning~\cite{masci2015geodesic,huang2019texturenet}, and so forth.
Since 3D data is usually acquired as human-designed CAD models and scans, manifold conversion from these data is an important problem.
However, these data is challenging to handle, especially because of the existence of sharp features, zero-volume structures (e.g. ShapeNet~\cite{chang2015shapenet}), and unavailability of correct exterior information (e.g. surface normal), which prevents existing methods from robustly producing high-quality watertight manifolds.
For example, the famous marching cubes algorithm~\cite{lorensen1987marching} requires exterior information to distinguish positive (exterior) from negative (interior) distances for zero-contour extraction based on a signed distance field.  However, there is no association of signs that successfully enables marching cubes to construct both the orange and green faces at the T-junction example shown in Figure~\ref{fig:intro-thin}(a).  Furthermore, the precision of marching cubes is only up to the side length of voxels, which makes it hard to preserve sharp features or sub-voxel structures in the final mesh.
Delaunay triangulation-based methods suffer from similar issues: To perfectly recover the geometry, both sides of the green edge in Figure~\ref{fig:intro-thin}(b) should be considered as exterior while only one side of the orange edge should be considered exterior. This natural ambiguity of exterior information is challenging to remove. Even with perfect exterior information, surface extraction step~\cite{boissonnat1984geometric} needs to keep some exterior volumes to avoid non-manifold edges at the T-junction (marked as the red spot) in Figure~\ref{fig:intro-thin}(b).

In this work, we target the problem of resolving exterior ambiguity and zero-volume structures and of obtaining a manifold that accurately resembles the input triangle soup.
As shown in Figure~\ref{fig:intro-thin}(c), we build an octree volume where regions surrounding the reference mesh are split into leaf nodes as voxels with a fine resolution. Instead of determining signed distances for these voxels and performing marching cubes, we simply mark voxels as occupied if they intersect the reference mesh.
Then, we determine exterior nodes as those who are connected to the boundary of the volume, without passing any occupied voxels.
We extract surfaces between exterior and occupied nodes to guarantee that generated surfaces are manifolds while preserving zero-volume structures.
There are two main advantages to this simple solution:
First, our method does not rely on the normal direction of faces in input meshes to decide exterior, as we can directly test whether one side of the face is connected to the exterior volume boundary. %, which is a prerequisite for Daulanay-based triangulation.
Second, we can preserve sub-voxel and zero-volume structures, as shown in Figure~\ref{fig:intro-thin}(c).
%by checking whether two neighboring voxels are both exterior and should be kept, resolving their orientation ambiguity. This also leads to our second advantage, that we robustly extract surfaces without missing zero-volume structures.

However, the vertices and faces on the extracted manifold with the aforementioned pipeline are always on the voxel grid, and cannot approximate the reference mesh geometry well.%, as face normals are non-smooth.
Therefore, we project extracted vertices to the nearest positions in the reference mesh (blue arrows in Figure~\ref{fig:intro-thin}(c)). However, doing this in a brute force way might cause problematic triangle inversions.
Our key technical contribution here is to formulate the projection as an energy minimization problem with inversion-free constraints and use a Gauss-Seidel strategy to solve the optimization efficiently.
Typically, a triangle inversion means that the triangle orientation is inconsistent with the orientation of its local neighborhood. Therefore, we can use vertex normal to describe the local orientation and detect triangle inversion if the dot product between the triangle normal and any of its vertices' normals is negative. 
Based on this observation, we propose to introduce three hard constraints for each triangle between its face normal and three vertex normals and optimize not only for vertex positions but also for vertex normals.
Since we want vertex positions to be close to the reference mesh and vertex normals to best describe local orientation, we design our energy as the sum of distances between vertices to the reference mesh and distances between vertex normals to the area-based interpolation of its incident face normals.
We observe that by solving for each variable while fixing all other variables, our proposed problem becomes convex. Therefore, we apply a Gauss-Seidel scheme and iteratively update each variable. Empirically, we find that the optimization converges with a small constant number of iterations. With additional sharp-preserving processing, we can detect and keep sharp features from the reference mesh.

We implement this algorithm and evaluate its performance based on several criteria.
For correctness, we verify whether our extracted mesh is topologically a watertight manifold and satisfies the inversion free condition.
For efficiency, we report the processing time for different complexity of input meshes.
For accuracy, we report the maximum and mean distance between vertices on the extracted mesh and the reference mesh.
Compared to previous state-of-the-art methods, ours is the only method that converts mesh to a watertight manifold correctly and efficiently with detailed feature preservation.
We demonstrate that our algorithm can handle zero-volume structures with a lot of self-intersections, organic shapes, and extremely complex meshes on a city scale.
Finally, with a small modification, our method can be used for a standard surface reconstruction purpose for scanning data.
% conclusion
In summary, our main contributions are:
 \begin{itemize}
 \setlength{\topsep}{0pt}
 \setlength{\parsep}{0pt}
 \setlength{\partopsep}{0pt}
 \setlength{\parskip}{0pt}
 \setlength{\itemsep}{2pt}
     \item Propose a novel surface extraction scheme that handles exterior ambiguity and zero-volume structures.
     \item Give a new formulation for mesh vertices projection optimization with triangle inversion-free constraints.
     \item Implement a robust and accurate software with thorough evaluations on large-scale datasets, with additional sharp-preservation features.
     \item Demonstrate our capability to handle complex meshes in large-scale and scanning data.
 \end{itemize}

\section{Related Work}
According to \cite{attene2013polygon}, existing methods can be classified as volumetric surface remeshing and direct local mesh repairing. Volumetric methods first convert the input triangle mesh into a volumetric representation surrounding the reference mesh, e.g., a sign distance field on voxels, and then reconstruct the manifold output from this volumetric representation using a contour extraction algorithm.  In this section, we discuss several popular volumetric representations, highlight our key difference from these methods in modeling the zero-contour, and briefly discuss local mesh repairing methods.  Finally, we compare our mesh projection process with other shape registration methods.

\paragraph{Volumetric Representation} Typical volumetric representations include regular grid~\cite{curless1996volumetric,nooruddin2003simplification} optionally accelerated with octrees~\cite{ju2004robust,hornung2006robust,kazhdan2006poisson,agarwala2007efficient,calakli2011ssd,kazhdan2013screened}, polyhedron from Delaunay triangulation~\cite{boissonnat1984geometric,chew1989constrained,shewchuk2002delaunay,rineau2007generic,hu2018tetwild} or BSP trees~\cite{murali1997consistent}. Although a polyhedron is better than a regular grid at keeping the exact input geometry, the constrained Delaunay triangulation is hard to implement to correctly handle all degenerate cases, e.g., non-zero volume structures, according to \cite{attene2018exact} . Furthermore, it takes a long time to subdivide the original mesh to avoid self-intersections as exact construction is required for robustness. Therefore, we choose the regular grid as our representation since it is easy to implement and does not suffer from precision or self-intersection problems, which makes it usually more scalable and robust. Although surface extracted from regular grids using marching cubes~\cite{lorensen1987marching,durst1988re,chernyaev1995marching,doi1991efficient} is not precise, we do not suffer from this issue since our newly proposed projection optimization moves extracted vertices precisely to the reference mesh.

\paragraph{Zero-contour Extraction} Zero-contour is commonly modeled with the help of a signed distance field. The sign of the distance is usually determined given surface normal~\cite{kazhdan2006poisson,agarwala2007efficient,calakli2011ssd,kazhdan2013screened}, scanning sensor location~\cite{curless1996volumetric,newcombe2011kinectfusion}, or geometry heuristics~\cite{boissonnat1984geometric,hornung2006robust}. One of the most geometrically meaningful and robust solutions is presented in~\cite{ju2004robust} where positive signs are determined by computing an occupancy grid and detecting exterior regions with a broad-first search from the volume boundary. Unfortunately, the aforementioned representations fail to model zero-volume structure as zero-contour since grids from both sides are associated with a positive sign. Instead, we do not compute the continuous signed distance for grid positions, but simply determine voxel status as occupied, exterior, or interior based on the occupancy grid. Zero-volume structures are therefore extracted as surfaces shared by occupied and exterior voxels. Further, while zero-contour determined by distance field with linear interpolation is inaccurate, our mesh projection optimization guarantees the accuracy in our final results.

\paragraph{Local Mesh Repairing.} These methods directly operate on the input mesh to avoid unnecessary change. A typical workflow is to address self-intersections by subdivisions~\cite{attene2014direct} with hybrid geometric kernel~\cite{attene2017imatistl}, remove non-manifold elements by singular vertex and edge decomposition~\cite{gueziec2001cutting,rossignac1999matchmaker} and greedily pair boundaries together. MeshFix~\cite{attene2010lightweight} works on solid objects and address problematic local regions. An integer optimization based on visual cues~\cite{chu2019repairing} can be introduced to achieve a more global mesh repairing strategy. Among them, MeshFix~\cite{attene2010lightweight} does not handle zero-volume structure. Other repairing methods can generate boundary edges during decomposition at T-junctions. These additional boundary edges are not ideal for geodesic analysis-related applications. As we test, \cite{attene2018exact} is quite robust for extracting high-quality watertight manifold by computing the outer hull. However, it introduces additional thickness for zero-volume structures and causes non-manifold issues in 0.8\% among our test shapes. Additionally, massive computation of self-intersections and triangulation makes this method less efficient. Comparing with mesh repairing methods, our method is efficient and robust and always guarantees a watertight manifold result.

\paragraph{Shape Registration} Our key challenge is to project our extracted mesh to the reference mesh. One related and widely researched problem is called shape registration, in which scenario a transformation function is optimized to register the source shape to the target using the iterative-closest-point (ICP) algorithm~\cite{besl1992method,chen1992object}. The transformation can be rigid~\cite{rusinkiewicz2001efficient,diez2015qualitative} or nonrigid~\cite{sumner2007embedded,li2008global,newcombe2015dynamicfusion}, and non-rigid ICP usually companies a rigidity regularization applied to the source mesh preserving local geometry features. Although we are handling the shape registration problem, we target at fixing the local geometry of our extracted mesh rather than preserving it. Therefore, common regularization energy does not apply. We propose to introduce hard inversion-free constraints rather than soft energy constraints for regularization. This gives maximum freedom for vertices to move within the constraints so that a nearly perfect fitting is possible. Although hard constraints introduce additional complexity, we derive a Gauss-Seidel optimization strategy to effectively solve this problem.
\section{Approach}
In this section, we discuss the details of our manifold conversion algorithm. Our input is a reference triangle mesh $\mathcal{M}_r = \{\mathcal{V}_r, \mathcal{F}_r\}$ with a set of vertices $\mathcal{V}_r$ and a set of triangles represented as vertex indices $\mathcal{F}_r$. We discuss surface extraction in Section~\ref{sec:approach-extract}, manifold optimization formulation in Section~\ref{sec:approach-optim}, our Gauss-Seidel solver in Section~\ref{sec:approach-solve}, and sharp preservation processing in Section~\ref{sec:approach-sharp}.

\subsection{Surface Extraction}
\label{sec:approach-extract}
We begin by constructing a set of voxels from $\mathcal{M}_r$ (e.g. orange voxels in Figure~\ref{fig:intro-thin}(c)) at a user-specified resolution using an octree representation. First, we normalize the reference mesh by re-centering it to the origin and applying a uniform scale so that all the absolute values of vertex coordinates are smaller than $1$. Next, we build the occupancy grid on octree with volume ranging from $-1.1$ to $1.1$. We assign the entire face set $\mathcal{F}_r$ to the root since it guarantees to contain the whole reference mesh. Then, we recursively split the octree until a certain node is not intersected by any triangles, or the maximum tree depth $H$ (specified by the user) is achieved. We mark nodes as occupied or empty depending on whether it intersects with any triangles. The overall time complexity for our octree construction is $O(|\mathcal{F}_r|\cdot H)$. The pseudo-code for building the octree can be found at Appendix~\ref{sec:appendix-octree}.

After we construct the octree, we build connections between all neighboring nodes that share faces. Since each node corresponds to a cube with six faces, we split neighbors into six connection groups, each of which corresponds to one of the node's faces. Note that each group may have multiple neighbors since neighboring octree nodes can have different resolutions. The pseudo-code of connection construction can be found at Appendix~\ref{sec:appendix-connection}.

\begin{figure}
\includegraphics[width=\linewidth]{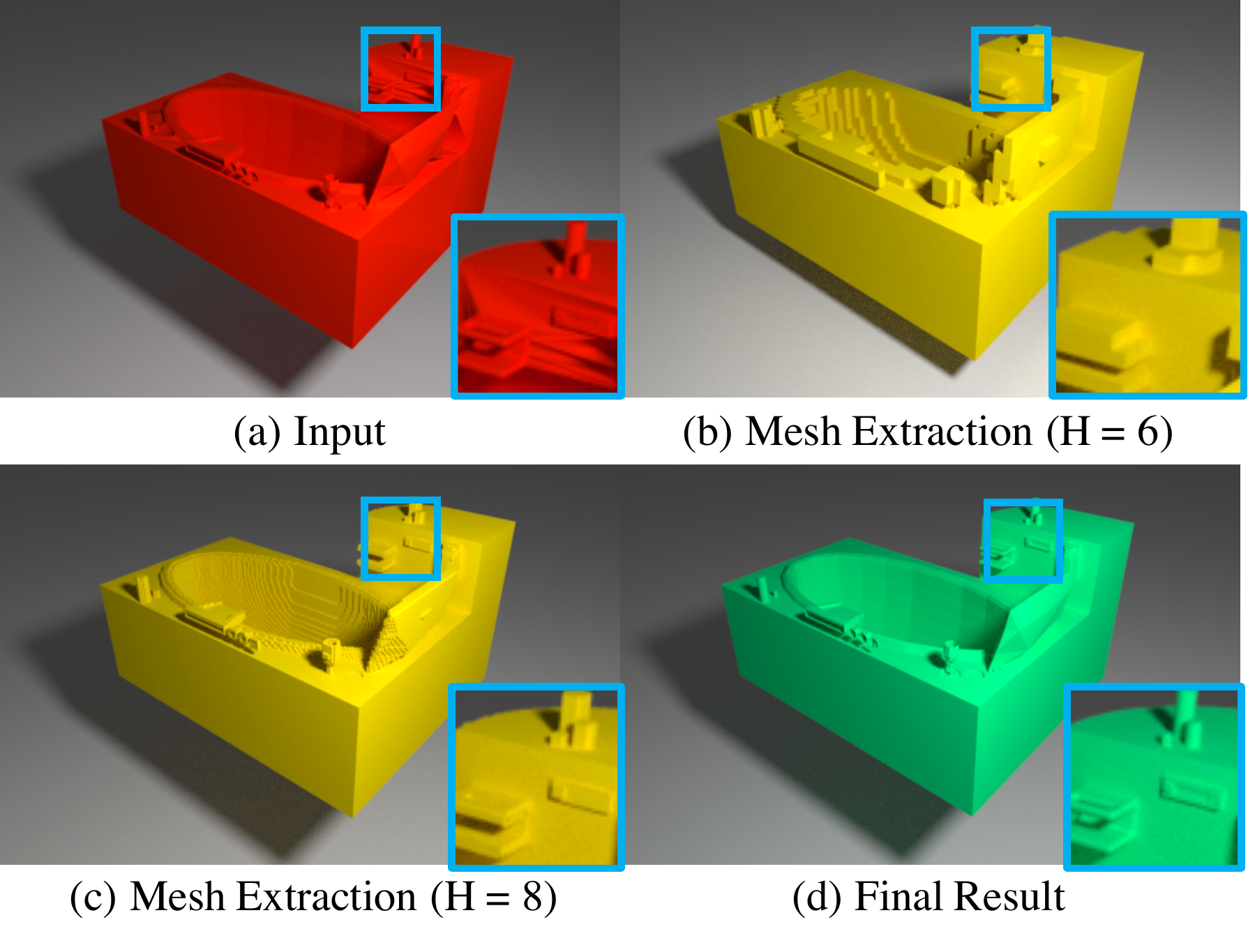}
\caption{Mesh extraction example. The input is in (a) where problematic face orientation causes rendering artifacts in the zoomed-in planar region. (b) and (c) are manifold meshes extracted using our method at different resolutions. (d) is our final result based on (c) followed by an optimization step (Section~\ref{sec:approach-optim}-\ref{sec:approach-sharp}), which is close to (a) but without less rendering artifacts.}
\label{fig:mesh-extraction}
\end{figure}

Boundary nodes can be easily detected as those with boundary faces whose corresponding connection groups have no neighbors.
Then, we treat leaf nodes and connections as a graph structure and apply a bread-first-search from boundary nodes until it reaches occupied nodes.
All visited empty nodes during the search are additionally marked as exterior nodes. We loop over each leaf occupied nodes and check whether any of its neighbors are marked as the exterior. We extract faces shared by occupied and exterior nodes as two triangles and collect all extracted triangles to form our new mesh.
%
%Note that the concept of boundary and exterior need be reinterpreted in the scenario of inside-out room scanning, where ``boundaries'' can be viewed as nodes covering any camera positions and ``exterior'' as nodes connecting to boundaries without going through occupied nodes.
Examples of extracted faces are shown in Figure~\ref{fig:mesh-extraction} (b) and (c).

\begin{figure}
\includegraphics[width=\linewidth]{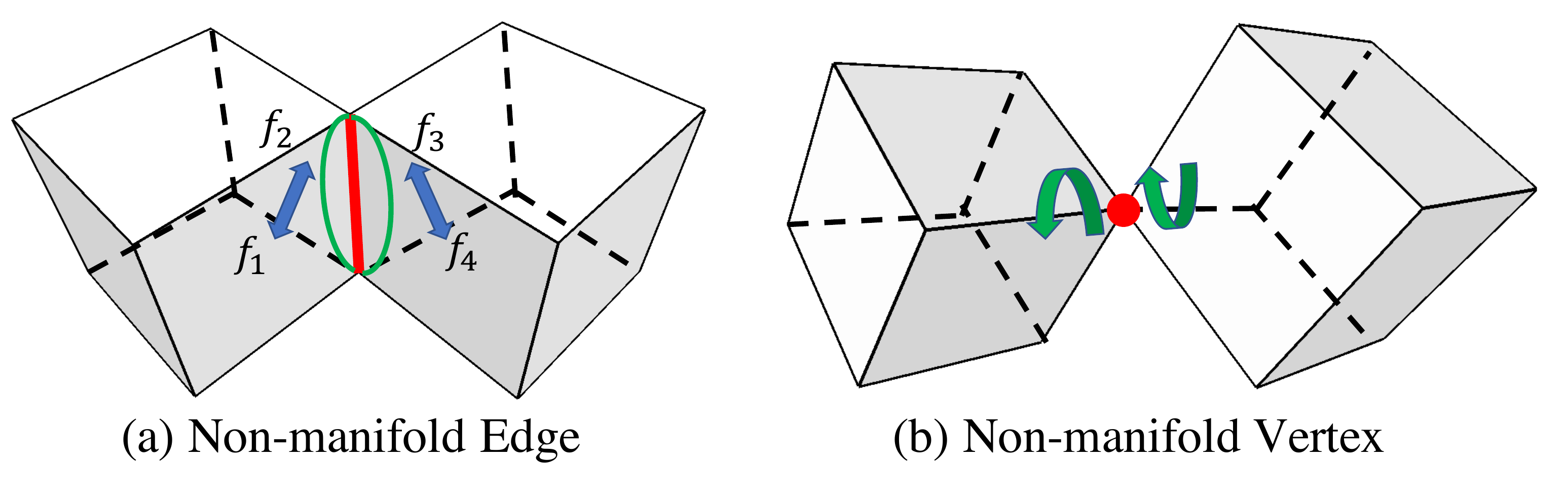}
\caption{Non-manifold edge and vertex examples. (a) Non-manifold edges (red) appears when two diagonal voxels are occupied among the edge's four incident voxels. (b) Non-manifold vertex is shared by more than one group of connected triangles.}
\label{fig:nonmanifold}
\end{figure}

During manifold extraction, there are two corner cases that prevent our extracted mesh from being a manifold.
First, non-manifold edges appear when exactly two diagonal voxels are occupied of the edge's four incident voxels, as shown in Figure~\ref{fig:nonmanifold}(a). We split the edge into two in order to disconnect the voxels.
In our implementation, we construct a standard half-edge structure where non-manifold edges are duplicated as four half-edges where each pair of half-edges belonging to the same voxel are marked as twins. For example, half-edges for face $f_1$ and $f_2$ are paired as a twin, $f_3$ and $f_4$ are paired as another twin in Figure~\ref{fig:nonmanifold}(a).
Second, a vertex could be shared by several groups of locally connected voxels, as shown in Figure~\ref{fig:nonmanifold}(b). We identify each separate group by traversing and collecting half-edges in the counter-clockwise order, split the vertex, and assign it to each group (green arrows in Figure~\ref{fig:nonmanifold}(b)).

After processing the above cases, we can guarantee that our extracted mesh is a perfect watertight manifold.

\subsection{Manifold Optimization Formulation}
\label{sec:approach-optim}
We aim at optimizing $\mathcal{M}_t$ and stick it to the reference mesh $\mathcal{M}_r$ to remove the voxel-shape artifacts. The energy of any vertex $k$ in the output mesh is defined as the squared distance to the nearest point $\mathbf{p}_k$ in the reference mesh:
\begin{equation}
E_D(\mathbf{v}_k) = \min_{\mathbf{p}_k \in \mathcal{M}_r} \|\mathbf{v}_k - \mathbf{p}_k\|_2^2,
\label{eq:distance}
\end{equation}
where $\mathbf{v}_k$ is a \emph{variable} representing the coordinate of vertex $k$ in the output.

In order to guarantee that the orientation of all the triangles is consistent with their neighborhood so that there is no flipping triangles, we associate each vertex $k$ with an \emph{variable} ${\mathbf{n}}_k \in \mathbb{R}^3$ representing its local orientation, i.e., the normal direction of that vertex. This leads to constraints that for each triangle $\Delta{abc}$ in the output, we have
\begin{equation}
\mathbf{n}_{abc} \cdot {\mathbf{n}}_k > 0 \;\;\;\forall k\in\{a,b,c\}.
\label{eq:consistent}
\end{equation}
Here, ${\mathbf{n}}_{abc}$ denotes the face normal of $\Delta{abc}$ on the output mesh computed from vertex position $\mathbf{v}_k$ (we denote face normal with 3 characters and vertex normals with 1 character in subscript).
Equation~\ref{eq:consistent} can be interpreted as the introduction of a vertex normal as a proxy to coordinate face orientations surrounding each vertex. To best represent its local orientation, we require vertex normals ${\mathbf{n}}_k$ to be as close to the interpolated normal direction $\tilde{\mathbf{n}}_k$ of vertex $k$:
\begin{equation}
E_N(\mathbf{n}_k,\tilde{\mathbf{n}}_k) = \|\mathbf{n}_k - \tilde{\mathbf{n}}_k\|^2_2
\label{eq:normal}
\end{equation}
The interpolated normal direction $\tilde{\mathbf{n}}_k$ is computed by averaging $\mathbf{n}_{abc}$ for all the triangles that are adjacent to vertex $k$.
%although more advanced algorithm like slerp interpolation can also apply.
Our problem formulation is summarized in the following optimization problem:
\begin{align}
\underset{\{\mathbf{v}_k\},\{\mathbf{n}_k\}}{\text{minimize}} \quad\;\;
&\sum_{k = 1}^{|\mathcal{V}_t|} E_D(\mathbf{v}_k)+ E_N(\mathbf{n}_k,\tilde{\mathbf{n}}_k)
\label{eq:optim}
\\
\text{subject to\;\;}\quad
& \mathbf{n}_{abc} \doteq(\mathbf{v}_b - \mathbf{v}_a) \times (\mathbf{v}_c - \mathbf{v}_a)
\nonumber\\
&\tilde{\mathbf{n}}_k \doteq \frac{\sum_{\Delta{abc}\in \mathcal{N}_k} \mathbf{n}_{abc}}{\left\lVert\sum_{\Delta{abc}\in \mathcal{N}_k} \mathbf{n}_{abc}\right\rVert_2}
\nonumber\\
& \mathbf{n}_{abc} \cdot \mathbf{n}_k > 0\;\;\;\;\;\;
\Delta abc \in \mathcal{F}_t, k\in\{a,b,c\}
\nonumber\\
& \|\mathbf{n}_k\|_2 = 1 \;\;\;\;\;\;\;\;\;\;
1 \leq k \leq |\mathcal{V}_t|.
\nonumber
\end{align}
Here, $\doteq$ is the symbol of ``defined as'' and $\mathcal{N}_k$ is the set of all triangles that are adjacent to vertex $k$.

\subsection{Solving the optimization}
\label{sec:approach-solve}
Equation~\ref{eq:optim} is non-trivial to solve since it is a large, non-convex system with complex constraints. However, we observe the following favored properties in the problem.
First, by initializing vertex positions $\mathbf{v}_k$ with the manifold generated by the method in Section~\ref{sec:approach-extract} and $\mathbf{n}_k$ as $\tilde{\mathbf{n}}_k$, all constraints in Equation~\ref{eq:optim} are satisfied because the initial mesh is inversion-free.
Second, this initialization is close to the optimal solution since the extracted mesh is not far from the reference mesh. This suggests that a proper local iterative method is promising to obtain decent results.
Third, we notice that by fixing all other variables and solve for a single one, the original problem can be converted to a convex optimization problem.
Therefore, we propose to apply a Gauss-Seidel method that iteratively updates the position and normal for each vertex fixing all other variables, where each update is a convex optimization.

\begin{figure}
\includegraphics[width=\linewidth]{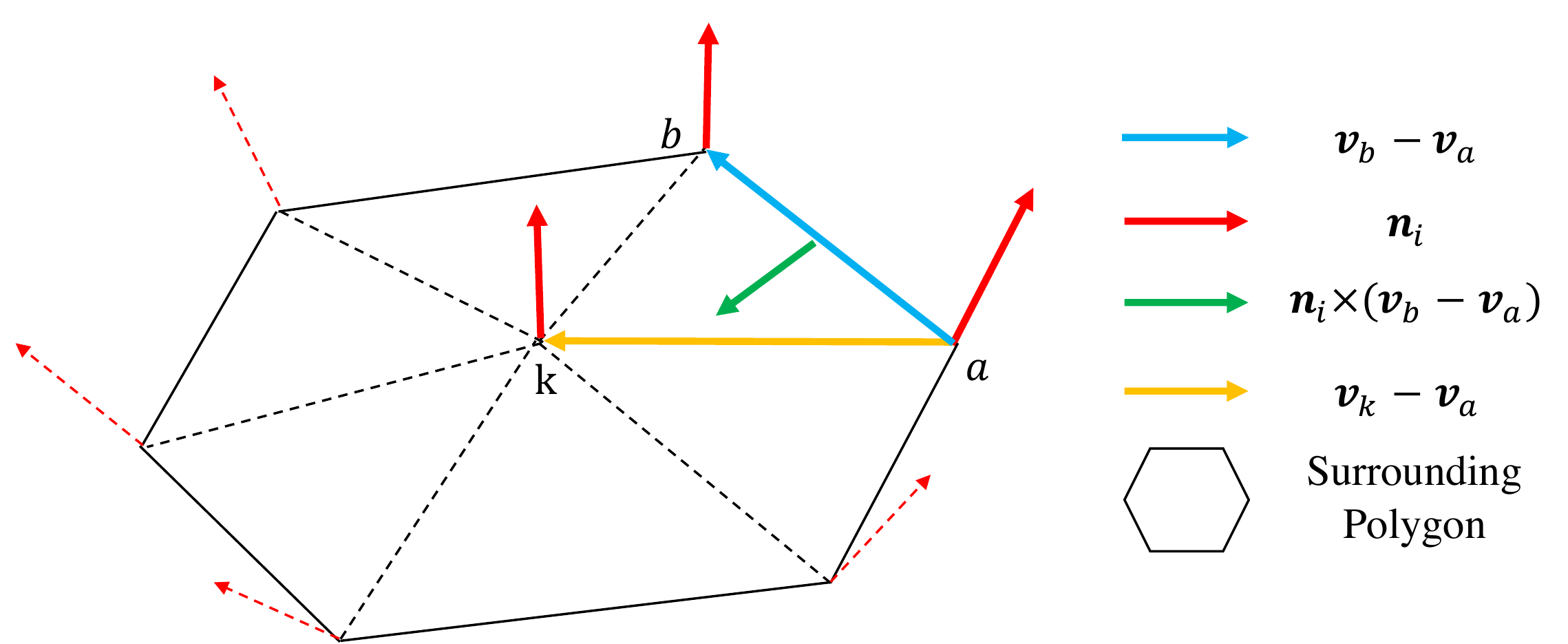}
\caption{Vertex $k$ are kept inside its surrounding polygon projected to local tangent planes specified by vertex normal of $i$ which equals or is adjacent to $k$.}
\label{fig:illus-vertex-update}
\end{figure}

\paragraph{Vertex Update} To update the position of $k$-th vertex, we find its nearest position $\mathbf{p}_k$ in the reference mesh $\mathcal{M}_r$, which can be efficiently computed by pre-storing an AABB tree~\cite{bergen1997efficient} for $\mathbf{M}_r$\footnote{\url{https://github.com/libigl/libigl/blob/master/include/igl/point_mesh_squared_distance.cpp}}. We aim at moving $\mathbf{v}_k$ to $\mathbf{p}_k$ maintaining all constraints. Since all face normals near vertex $k$ are changed during the vertex movement, consistency constraints can be violated between face normals for any of these triangles and their vertices' normals. Therefore, a subproblem can be extracted from Equation~\ref{eq:optim} as Equation~\ref{eq:vertex-update}.
We set $\varepsilon_v=10^{-5}$ as a positive threshold
\begin{align}
\underset{\mathbf{v}_k}{\text{minimize}}\;\;& \|\mathbf{v}_k - \mathbf{p}_k\|_2^2
\label{eq:vertex-update}\\
\text{subject to\;\;}& \mathbf{n}_{kab}\cdot \mathbf{n}_i \ge \varepsilon_v
\nonumber\\
& \forall\Delta_{kab}\in \mathcal{F}_t,\;
i\in\{k,a,b\}
\nonumber
\end{align}
Here, $\mathbf{n}_{kab} = (\mathbf{v}_a - \mathbf{v}_k) \times (\mathbf{v}_b-\mathbf{v}_k)$ is linear with respect to $\mathbf{v}_k$ as $\mathbf{v}_k \times \mathbf{v}_k = \mathbf{0}$.
As shown in Figure~\ref{fig:illus-vertex-update}, this linear relationship is also geometrically meaningful: we require vertex $\mathbf{v}_k$ to sit inside its surrounding 2D polygon by projecting the neighborhoods to the local tangent planes specified by any vertex normal which is or is adjacent to $k$.
%$(\mathbf{v}_b-\mathbf{v}_a)\times(\mathbf{v}_k-\mathbf{v}_a)\cdot\mathbf{n}_i=\mathbf{n}_i\times(\mathbf{v}_b-\mathbf{v}_a)\cdot(\mathbf{v}_k-\mathbf{v}_a)$, where $\mathbf{n}_i\times(\mathbf{v}_b-\mathbf{v}_a)$ (Figure~\ref{fig:illus-vertex-update} green arrow) represents the left-hand-side direction of edge $\left<\mathbf{v}_a,\mathbf{v}_b\right>$ (Figure~\ref{fig:illus-vertex-update} blue arrow) in the local tangent plane specified by $\mathbf{n}_i$ (Figure~\ref{fig:illus-vertex-update} solid red arrow). Positive value of the dot product between this red arrow and $\mathbf{v}_k-\mathbf{v}_a$ (Figure~\ref{fig:illus-vertex-update} orange arrow) implies that we require $\mathbf{v}_k$ to sit at the left-hand side of $\left<\mathbf{v}_a,\mathbf{v}_b\right>$. Collecting all constraints,  In other word, the linearization of normal constraints leads to another intuitively geometric interpretation of inversion-free condition from the perspective of vertex positions.
Therefore, Equation~\ref{eq:vertex-update} is a convex problem with quadratic energy and linear constraints. We solve it with the simplex algorithm~\cite{vanderbei2015linear}: Initially, we set the target position as $\mathbf{p}_k$. First, we move $\mathbf{v}_k$ towards the target until it reaches certain boundary constraint. Second, we project the target position to set of activated boundary constraints and go back to the first step. The algorithm terminates when we reach the target position or trapped into a corner when the rank of boundary constraints is 3. Since each constraint is becoming the boundary for at most once, the time complexity is linear to the number of constraints. The amortized time complexity over the whole mesh for vertex updates is $O(|\mathcal{F}_t|)$.

\paragraph{Normal Update}
To update $k$-th vertex normal $\mathbf{v}_k$, we first compute the target normal $\tilde{\mathbf{n}}_k$ by interpolating the current neighboring face normals. We extract the related energy term and constraints from Equation~\ref{eq:optim} and solve Equation~\ref{eq:normal-update}, where we set $\varepsilon_n=10^{-2
}$.
\begin{align}
\underset{\mathbf{n}_k}{\text{minimize}}\;\;& \|\mathbf{n}_k - \tilde{\mathbf{n}}_k\|_2^2
\label{eq:normal-update}\\
\text{subject to\;\;}& \mathbf{n}_{kab}\cdot \mathbf{n}_k \ge \varepsilon_n
\quad\quad\Delta_{kab}\in \mathcal{F}_t
\nonumber\\
& \|\mathbf{n}_k\|_2 = 1
\nonumber
\end{align}
Although the unit-vector constraint $\|\mathbf{n}_k\|_2 = 1$ makes this problem non-convex, we can relax the problem by removing it and solve a convex problem with solution denoted as $\hat{\mathbf{n}}_k$. Note that $\mathbf{n}_{kab}$ are computed from vertex positions, which are considered as constants in the normal update process. Then, we derive Theorem~\ref{th:norm}.
\begin{theorem}
The global optimal solution of Equation~\ref{eq:normal-update} is identical to $\frac{\hat{\mathbf{n}}_k}{\|\hat{\mathbf{n}}_k\|}$, where $\hat{\mathbf{n}}_k$ is the global optimal solution of the same problem but removing the constraint $\|\mathbf{n}_k\|_2 = 1$.
\label{th:norm}
\end{theorem}
\begin{figure}
\includegraphics[width=\linewidth]{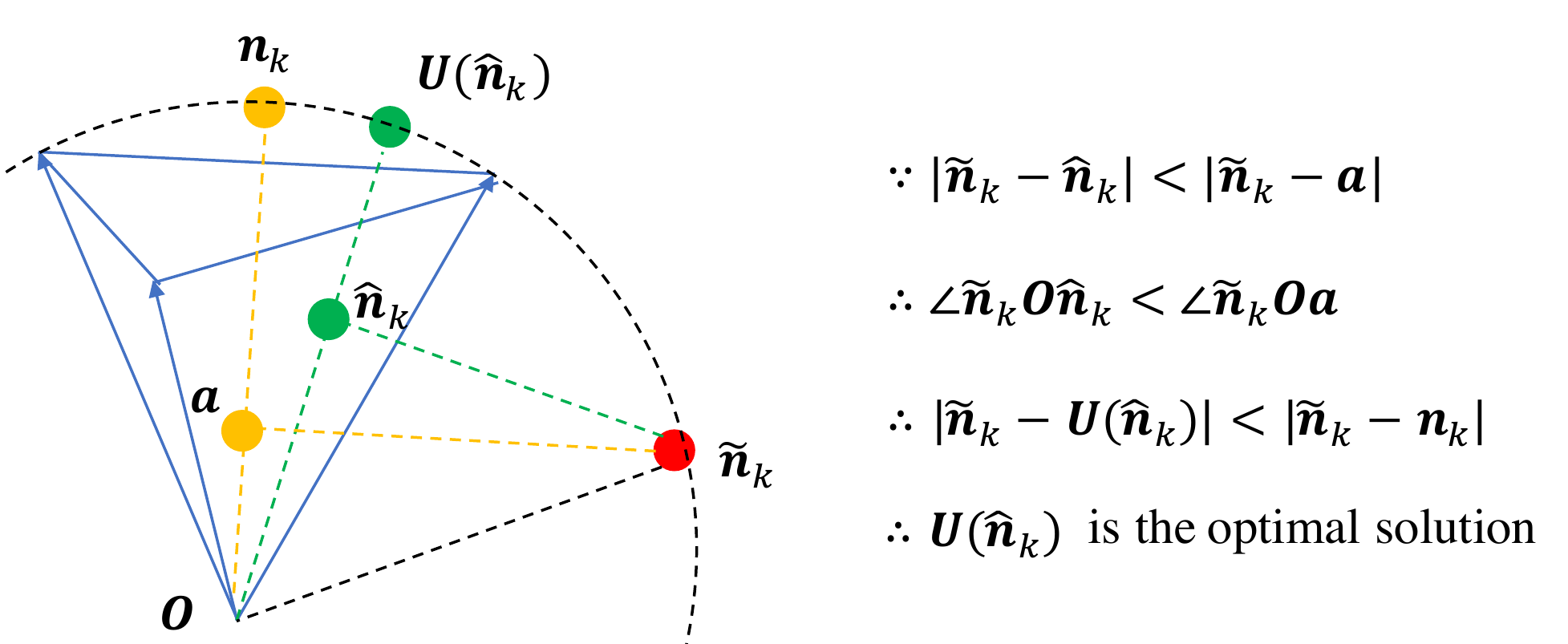}
\caption{Illustration of optimal normal solution for Equation~\ref{eq:normal-update}. Feasible solutions sits on the sphere surface inside the 3D polygon specified by linear constraints. Optimal solution is the normalization of the point $\hat{\mathbf{n}}_k$ inside the polygon that is closest to $\tilde{\mathbf{n}}_k$.}
\label{fig:proof}
\end{figure}
\begin{proof}
If $\tilde{\mathbf{n}}_k$ satisfies all constraints, the theorem is valid because the energy is zero.

Otherwise, $\tilde{\mathbf{n}}_k$ is outside the volume specified by the constraints as a polygon shown in Figure~\ref{fig:proof} (blue lines). $\hat{\mathbf{n}}_k$ is the projection of $\tilde{\mathbf{n}}_k$ to the boundary of the polygon (green spot inside the sphere) and $\frac{\hat{\mathbf{n}}_k}{\|\hat{\mathbf{n}}_k\|}$ is shown as the green spot at the sphere boundary. For any other feasible solution $\mathbf{n}_k$, we can project $\tilde{\mathbf{n}}_k$ to line $\left<\mathbf{O},\mathbf{n}_k\right>$ as $\mathbf{a}$. Since $\hat{\mathbf{n}}_k$ is the optimal solution without unit-vector constraints, it is closer to $\tilde{\mathbf{n}}_k$ than $\mathbf{a}$. Therefore, the the angle between $\hat{\mathbf{n}}_k$ and $\tilde{\mathbf{n}}_k$ is smaller than that between $\mathbf{a}$ and $\tilde{\mathbf{n}}_k$. Therefore, $\frac{\hat{\mathbf{n}}_k}{\|\hat{\mathbf{n}}_k\|}$ is closer to any other solution $\mathbf{n}_k$, and is the global optimal for Equation~\ref{eq:normal-update}.
\end{proof}

According to Theorem~\ref{th:norm}, we update vertex normal with $\mathbf{n}_k \gets \frac{\hat{\mathbf{n}}_k}{\|\hat{\mathbf{n}}_k\|}$ as the optimal solution.  Note that $\hat{\mathbf{n}}_k$ can be solved using the same algorithm applied to Equation~\ref{eq:vertex-update}. Therefore, the time complexity for updating all vertex normals in one pass is also $O(|\mathcal{V}_t|)$.

\paragraph{Gauss-Seidel Update} We maintain an active vertex list before each pass of vertex and normal update. During each pass, we loop over the vertex list in the decreasing order of $E_D(\mathbf{v}_k)$, update the vertex position by solving Equation~\ref{eq:vertex-update} and vertex normal by solving Equation~\ref{eq:normal-update}. If a vertex is updated, the vertex and its adjacent vertices are inserted into the active vertex list for the next pass. We terminate the algorithm when no more vertex is updated. Initially, the active vertex list contains all vertex in $\mathcal{V}_t$. In practice, we find that the total number of vertex updates is at the scale of $O(|\mathcal{F}_t|)$ with a constant coefficient smaller than 10.  Therefore, our algorithm is efficient for handling meshes on a large scale.

\subsection{Sharp Preservation}
\label{sec:approach-sharp}
\begin{figure}
\includegraphics[width=\linewidth]{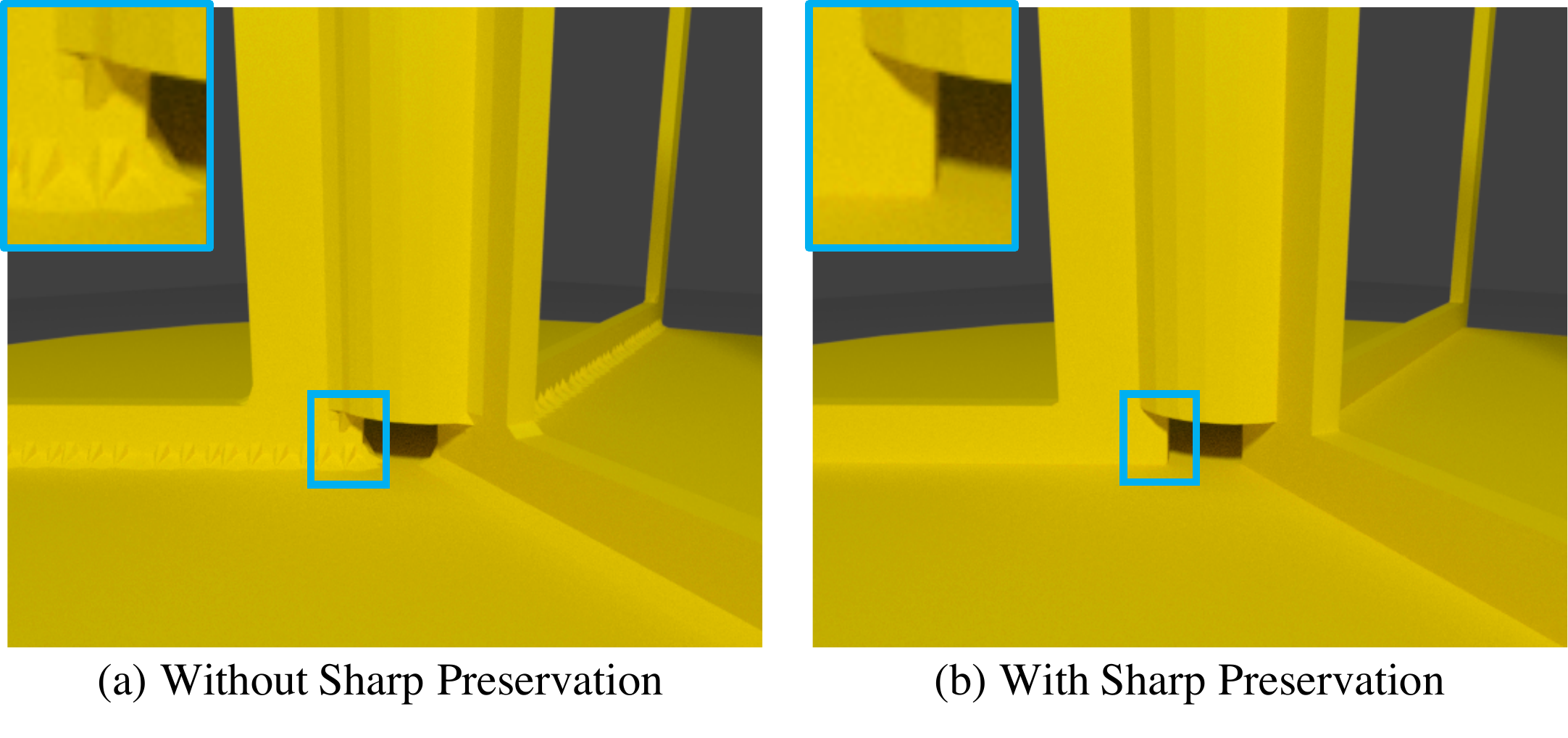}
\caption{Sharp feature preservation example. By projecting newly added vertices to sharp locations, sharp features can be preserved.}
\label{fig:sharp}
\end{figure}

Although $\mathcal{M}_t$ is stitched to the reference mesh $\mathcal{M}_r$ during optimization at mesh vertices $\mathcal{V}_t$, there is no guarantee that edges also perfectly sit in $\mathcal{M}_r$. This leads to the failure of sharp feature preservation as shown in Figure~\ref{fig:sharp}(a). However, it also gives us intuition to detect problematic edges: an edge breaks the sharp feature if its midpoint is not close enough ($10^{-3}$ of voxel size in our implementation) to $\mathcal{M}_r$. Therefore, we cut such an edge at its midpoint and move it to the sharp feature line in $\mathcal{M}_r$ using the same vertex and normal update in Section~\ref{sec:approach-solve}.
\begin{figure}
\includegraphics[width=\linewidth]{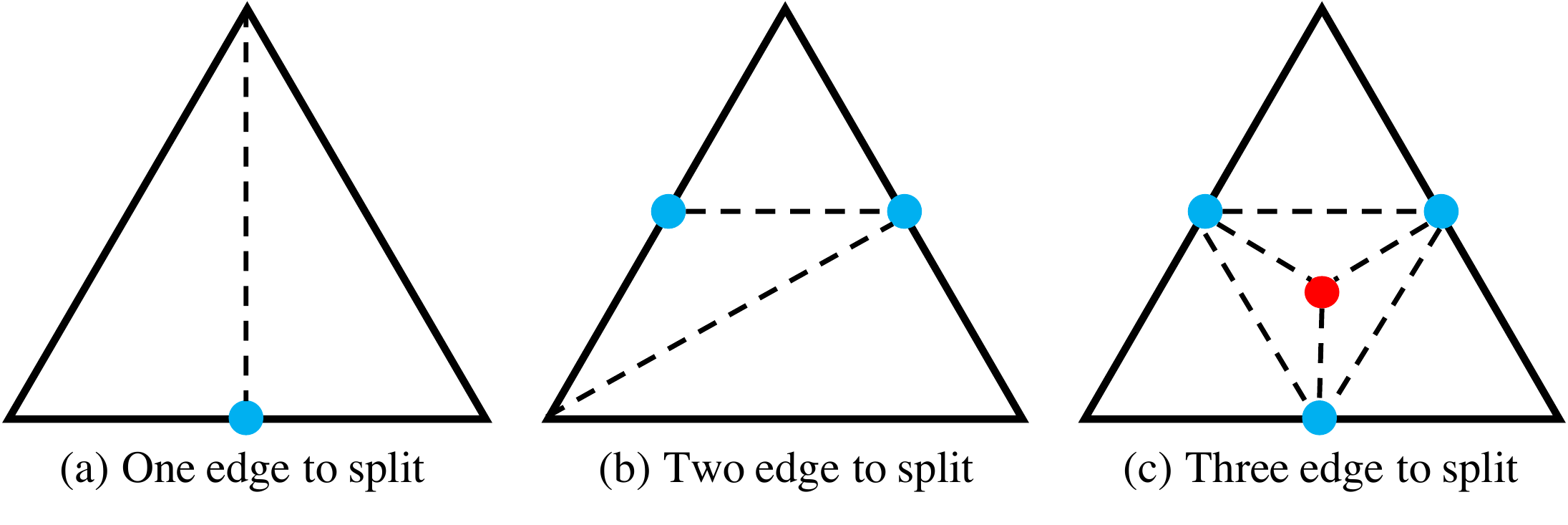}
\caption{We subdivide a triangle based on the number of edges to cut. Blue vertices are moved to sharp edges and red vertices are moved to sharp corners.}
\label{fig:cut}
\end{figure}
We first detect all edges that need to be cut, and subdivide the triangles depending on the number of edges to cut as shown in Figure~\ref{fig:cut}, where we aim to move blue vertices to sharp edges and the red vertex is to the sharp corner. For an edge with blue vertex, the target position is the projection of the midpoint to the intersection line of planes in the reference mesh where the edge endpoints sit in. For the triangle with the red vertex, the corner is determined by intersecting three planes of the reference mesh where the triangle vertices sit in. Figure~\ref{fig:sharp} demonstrates the before and after using sharp preserving processing.
\begin{figure*}
\begin{minipage}{0.33\linewidth}
\centering
\includegraphics[width=\linewidth]{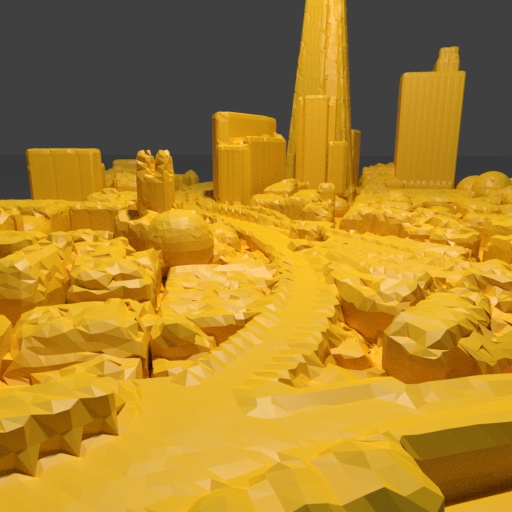}
Manifold
\end{minipage}
\begin{minipage}{0.33\linewidth}
\centering
\includegraphics[width=\linewidth]{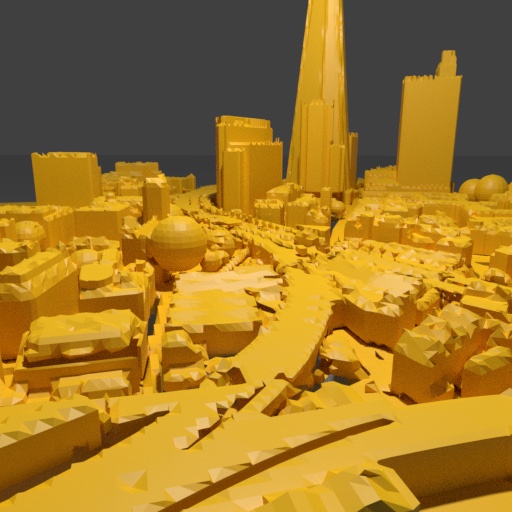}
PolyMender
\end{minipage}
\begin{minipage}{0.33\linewidth}
\centering
\includegraphics[width=\linewidth]{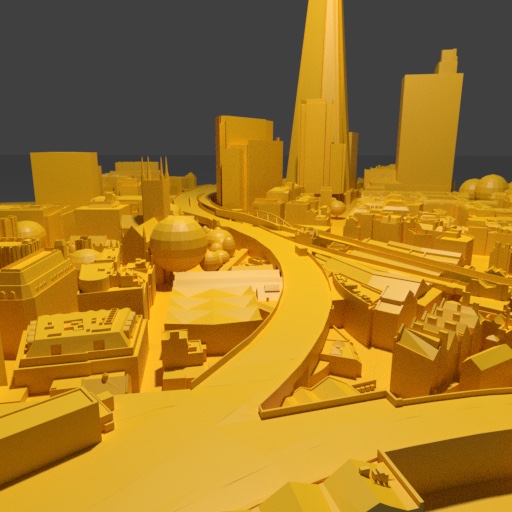}
Ours
\end{minipage}
\caption{City manifold results successfully processed by different methods. Our method is the only one that preserves details of the city.}
\label{fig:visual-city}
\end{figure*}
\section{Results}
\begin{figure*}
\includegraphics[width=\linewidth]{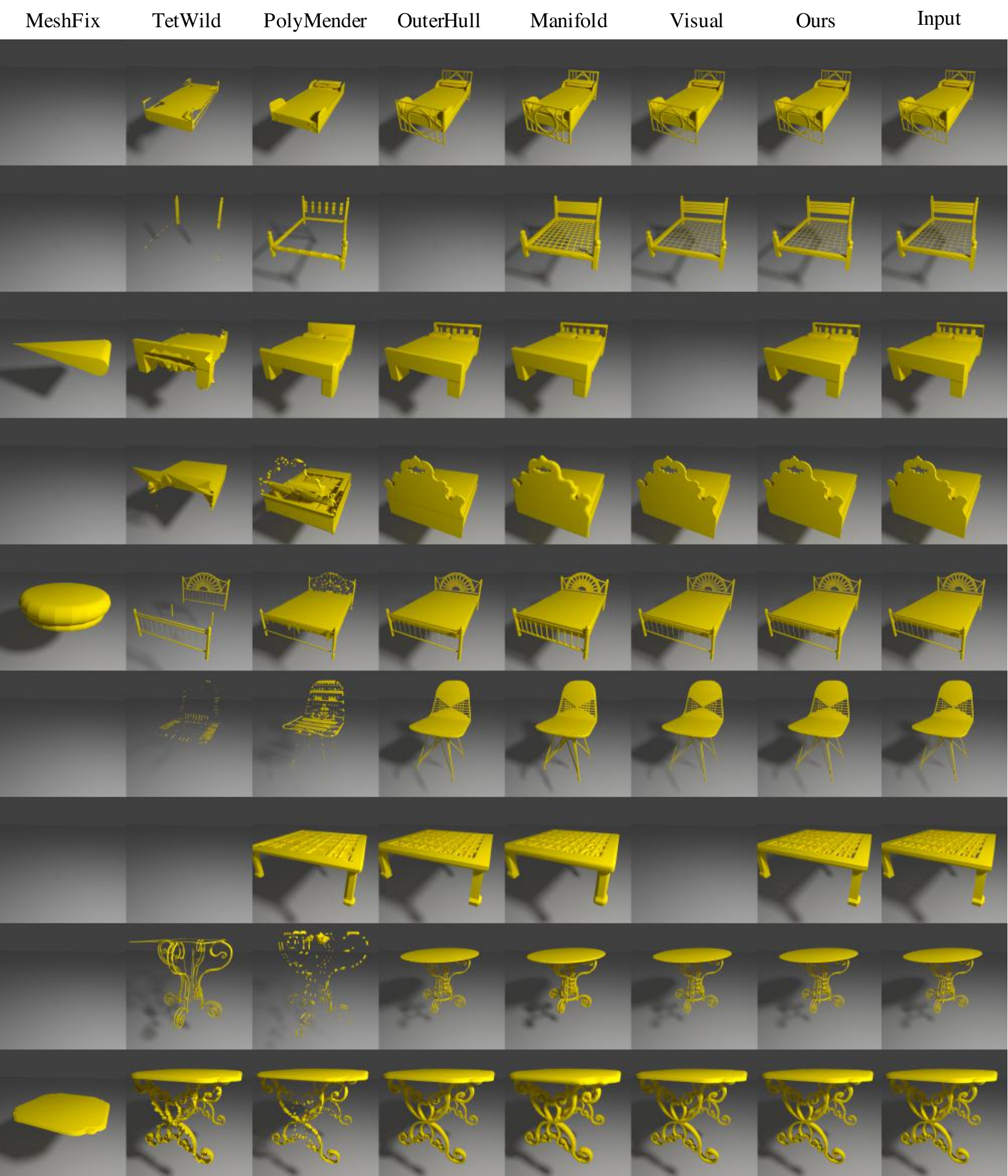}
\caption{Visual results generated from different methods on ModelNet10~\cite{wu20153d}.}
\label{fig:correctness}
\end{figure*}

We evaluate the performance of our algorithm based on several criteria related to correctness, efficiency and accuracy. We compare with several state-of-the-art methods including
``MeshFix''~\cite{attene2010lightweight},
``TetWild''~\cite{hu2018tetwild},
``PolyMender''~\cite{ju2004robust},
``OuterHull''~\cite{attene2017imatistl,attene2018exact},
``Manifold''~\cite{huang2018robust}, and ``Visual''~\cite{chu2019repairing}.
We massively evaluate proposed methods for all models from the ModelNet10 dataset~\cite{wu20153d}. Visual results are shown in Figure~\ref{fig:correctness}.

\paragraph{Correctness.} We evaluate whether a result is topologically an oriented watertight 2-manifold by checking the existence of boundary edges, non-manifold edges (NM Edges), non-manfiold vertices (NM Vertices), and triangle inversions based on our proposed criterion. Table~\ref{tab:correct} shows the comparison. As a result, MeshFix~\cite{attene2010lightweight}, TetWild~\cite{hu2018tetwild} and Visual~\cite{chu2019repairing} fail to process hundreds of shapes 
for ModelNet10. PolyMender~\cite{ju2004robust}, Manifold~\cite{huang2018robust} and our method are robust to process all shapes. We consider OuterHull~\cite{attene2017imatistl,attene2018exact} as almost robust since it fails to process less than 1\% among all shapes. Among all successfully-processed models, 38 from OuterHull have non-manifold edges or vertices. Although both Manifold~\cite{huang2018robust} and Visual~\cite{chu2019repairing} are free from non-manifold edges, Manifold~\cite{huang2018robust} yields non-manifold vertices and Visual~\cite{chu2019repairing} yields boundary edges. Our method robustly handles these problems correctly and generates results that capture geometry details in the input as shown in Figure~\ref{fig:correctness}.
\begin{table}
\begin{tabular}{|c|c|c|c|c|}
\hline
& Boundary & NM Edges & NM Vertices & Failure\\
\hline
MeshFix & \textbf{0} & \textbf{0} & \textbf{0} & 2416\\
\hline
TetWild & \textbf{0} & 2465 & 2545 & 1421\\
\hline
PolyMender & \textbf{0} & \textbf{0} & \textbf{0} & \textbf{0}\\
\hline
OuterHull & 38 & \textbf{0} & 38 & 37\\
\hline
Manifold & \textbf{0} & \textbf{0} & 210 & \textbf{0} \\
\hline
Visual & 3250 & \textbf{0} & 3259 & 406 \\
\hline
Ours & \textbf{0} & \textbf{0} & \textbf{0} & \textbf{0}\\
\hline
\end{tabular}
\caption{Number of results violating certain criteria in ModelNet10 among 4899 models.}
\label{tab:correct}
\end{table}

\paragraph{Efficiency.} We report the processing time of different methods on different input shapes. As shown in Table~\ref{tab:time}. PolyMender~\cite{ju2004robust}, Manifold~\cite{huang2018robust}, and our method can be considered as efficient since these methods can process object-level models (Bathtub, Gargoyle, and Rampant) in less than 10 seconds. Additionally, only these three methods can handle the city-scale CAD model in our experiment. The city results from these methods are visualized in Figure~\ref{fig:visual-city}. Among these methods, ours is the only one that preserves details of the entire city.

\begin{table}
\begin{tabular}{|c|c|c|c|c|}
\hline
& Bathtab & Gargoyle & Rampant & HoliCity\\
\hline
MeshFix & 68.8 & 2.0 & 2.4 & Fail\\
\hline
TetWild & 89.7 & 733 & 238 & Fail\\
\hline
PolyMender & 2.6 & 4.3 & 2.3 & 19.7\\
\hline
OuterHull & 65.5 & 5.2 & 5.7 & Fail\\
\hline
Manifold & 2.4 & 3.3 & 4.2 & 28.7 \\
\hline
Visual & Fail & 20.2 & 39.5 & Fail\\
\hline
Ours & 3.7 & 5.2 & 4.6 & 271\\
\hline
\end{tabular}
\caption{Time consumed for different methods to process different input meshes.}
\label{tab:time}
\end{table}

\paragraph{Accuracy.} We measure the remeshing accuracy by the maximum and mean distance between extracted vertices to the reference mesh, denoted as T2R-max and T2R-mean. In order to demonstrate the capability of zero-volume structure preservation, we additionally evaluate the coverage by measuring the maximum and mean distance from exterior regions of the reference mesh to the reconstructed mesh, denoted as R2T-max and R2T-mean. We compute the average of these terms for all processed models with different methods, and report scores in Table~\ref{tab:score}. For this experiment, we normalize each object with a uniform scale so that the maximum axis length of its bounding box is two units.
\begin{table}
\begin{tabular}{|c|c|c|c|c|}
\hline
& T2R-max & T2R-mean & R2T-max & R2T-mean\\
\hline
MeshFix & \textbf{0} & \textbf{0} & 1.9$\times 10^{-1}$ & 5.0$\times 10^{-2}$\\
\hline
TetWild & 2.7$\times 10^{-2}$ & 9.3$\times 10^{-4}$ & 2.1$\times 10^{-1}$ & 6.3$\times 10^{-2}$\\
\hline
PolyMender & 3.8$\times 10^{-2}$ & 2.2$\times 10^{-3}$ & 6.3$\times 10^{-2}$ & 6.2$\times 10^{-3}$\\
\hline
OuterHull & 4.6$\times 10^{-3}$ & 3.8$\times 10^{-4}$ & 4.5$\times 10^{-3}$ & 3.3$\times 10^{-4}$\\
\hline
Manifold & 1.2$\times 10^{-2}$ & 7.6$\times 10^{-3}$ & 3.1$\times 10^{-2}$ & 5.4$\times 10^{-3}$ \\
\hline
Visual & 3.2$\times 10^{-3}$ & 1.1$\times 10^{-4}$ & 8.9$\times 10^{-3}$ & 1.9$\times 10^{-4}$ \\
\hline
Ours & 1.2$\times 10^{-3}$ & 8.9$\times 10^{-6}$ & \textbf{3.3$\mathbf{\times 10^{-3}}$} & \textbf{7.3$\mathbf{\times 10^{-6}}$}\\
\hline
\end{tabular}
\caption{Fitting quality between the exterior regions of input mesh and results from different methods.}
\label{tab:score}
\end{table}
Although direct comparison is not valid since different methods successfully handle different sets of objects, we can draw some conclusions from the scale of the errors. MeshFix~\cite{attene2010lightweight}, TetWild~\cite{hu2018tetwild} and PolyMender~\cite{ju2004robust} cannot preserve the original geometry well where maximum geometry errors is at the scale of $10^{-2}$. Manifold~\cite{huang2018robust} preserves better geometry but tends to produce thicker geometry and cannot preserve sharp features. OuterHull~\cite{attene2017imatistl,attene2018exact}, Visual~\cite{chu2019repairing} and our method nearly preserve all geometry details, and our method achieves smallest fitting error. The above discussed problems are reflected in Figure~\ref{fig:correctness}. Among all succesfully generated models, OuterHull~\cite{attene2017imatistl,attene2018exact}, Visual~\cite{chu2019repairing} and our method can produce results quite similar to the input CAD models.

\begin{figure}
\includegraphics[width=\linewidth]{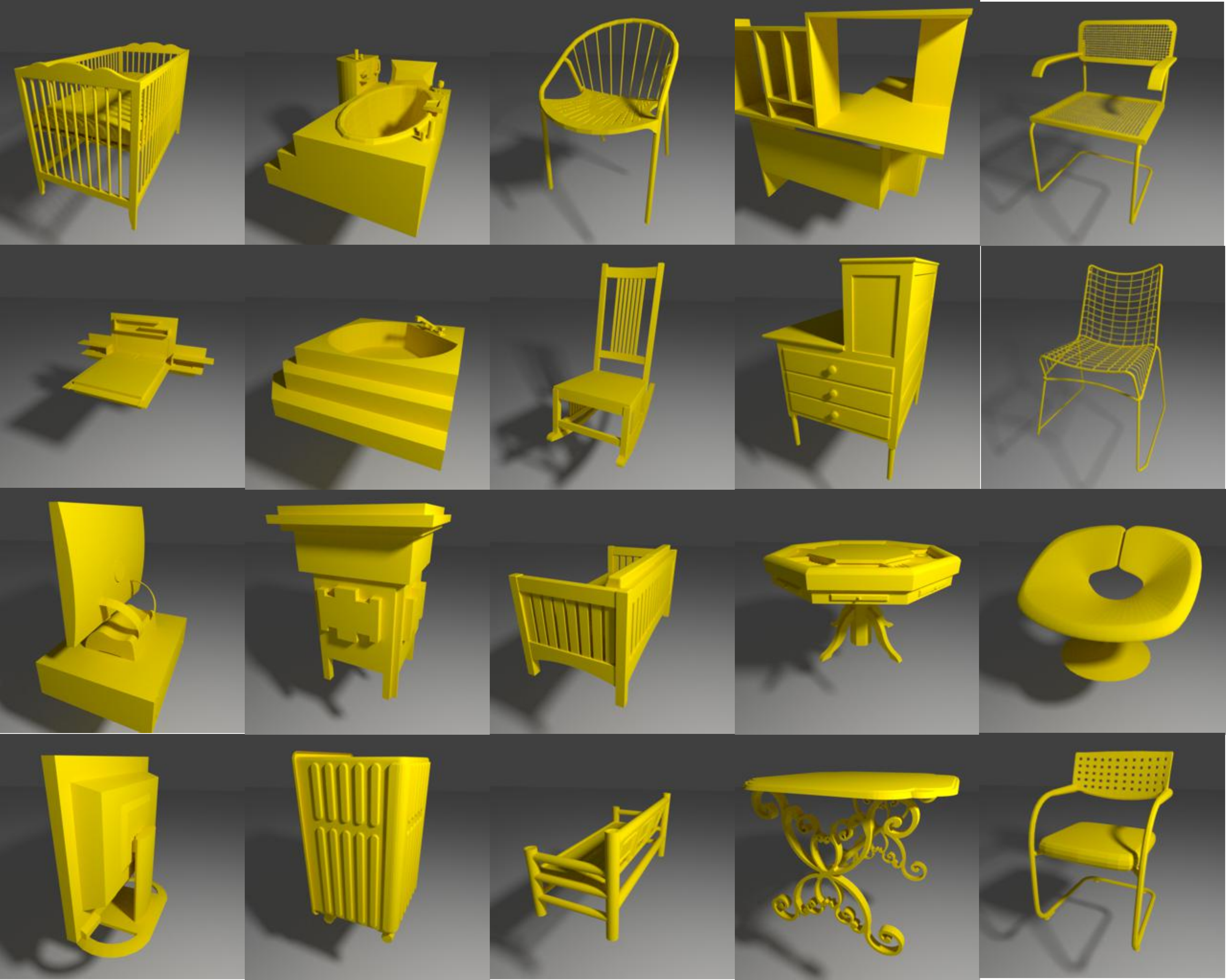}
\caption{Visualization of ModelNet10 manifold result generated by our method.}
\label{fig:result}
\end{figure}

\begin{figure}
\includegraphics[width=\linewidth]{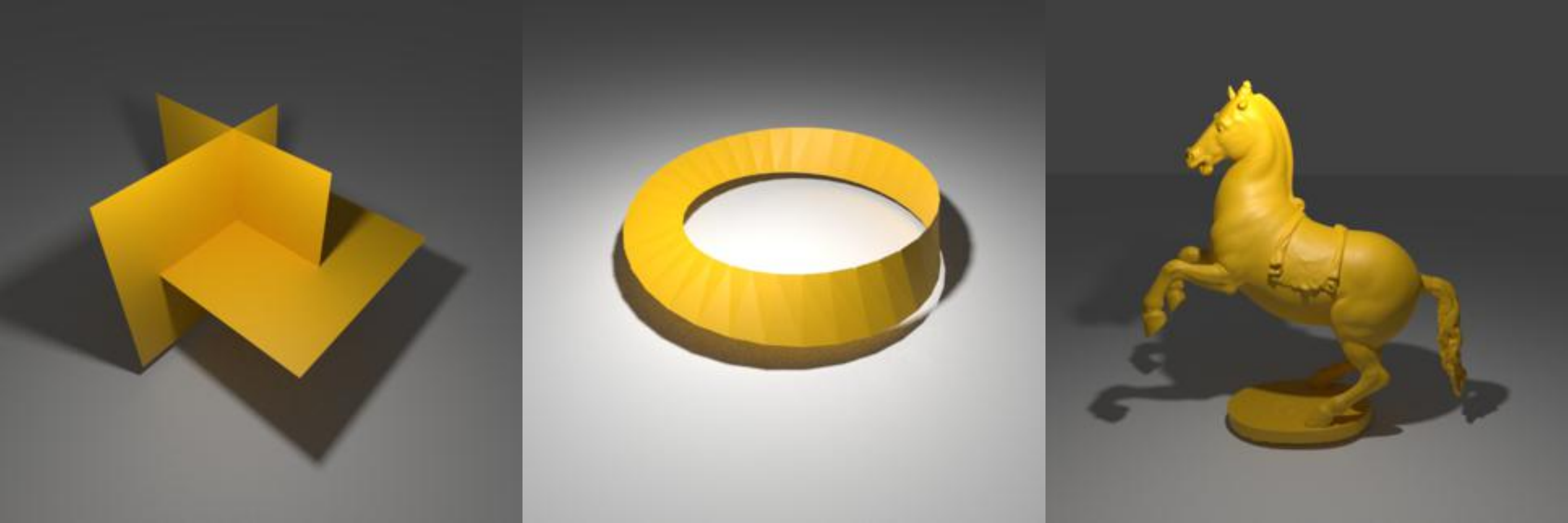}
\caption{We successfully create manifold for models with self-intersections, and can deal with m$\ddot{o}$bius strip and organic shapes.}
\label{fig:interesting-result}
\end{figure}

\paragraph{Remeshing Triangles} In Figure~\ref{fig:result}, we visualize various manifold models that we produce in ModelNet10 with faithful geometry detail preservation. In these examples, we are able to preserve sharp features, zero-volume structures and tiny holes while guarantee watertight manifold topology.
Additionally, we successfully create manifold for models with self-intersections, and can deal with m$\ddot{o}$bius strip or organic shapes, as shown in Figure~\ref{fig:interesting-result}.

\begin{figure}
\includegraphics[width=\linewidth]{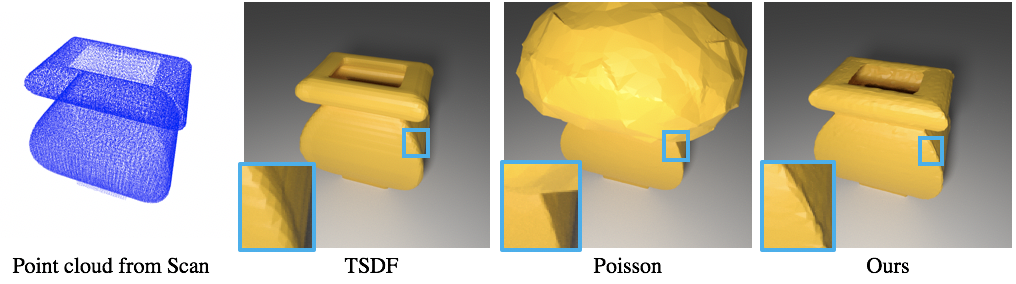}
\caption{Surface reconstruction from range images. Our method provides more accurate reconstruction and is robust to large holes in scans.}
\label{fig:recon}
\end{figure}

\paragraph{Remeshing Scan} We slightly change our algorithm to enable surface reconstruction from scanning data. Our input is aligned multiview depth images and our output is a reconstructed triangle mesh. We simply extract point clouds from all range images, merge them together in the world coordinate system, and run our algorithm to obtain the reconstruction, where the only difference is to compute $E_D$ based on nearest point-to-point distance instead of point-to-mesh distance. We compare our algorithm with commonly used Poisson surface reconstruction~\cite{kazhdan2006poisson} and TSDF reconstruction~\cite{curless1996volumetric}. By Poisson reconstruction, we compute the point normals of the extracted point cloud based on jet fitting and apply the Poisson surface reconstruction implemented in CGAL. Table~\ref{tab:recon} reports the chamfer distances in both ways between the ground truth and results generated from different methods. Visualization of surface reconstruction is shown in Figure~\ref{fig:recon}.
\begin{table}
\begin{tabular}{|c|c|c|c|}
\hline
& Poisson & TSDF & Ours\\
\hline
Point-to-Scan & 5.44$\times 10^{-2}$ & 2.71$\times 10^{-2}$ & \textbf{2.56$\mathbf{\times 10^{-2}}$} \\
\hline
Scan-to-Point & 23.6$\times 10^{-2}$ & 8.73$\times 10^{-2}$ & \textbf{1.48$\mathbf{\times 10^{-2}}$} \\
\hline
\end{tabular}
\caption{Chamfer distance between the ground truth point cloud and the surface reconstruction result.}
\label{tab:recon}
\end{table}
TSDF reconstruction suffers from inaccurate per-frame distance field estimation, which causes relatively large errors at the edges (Figure~\ref{fig:recon}. Poisson reconstruction is suffering from large holes where there is no observed data. Distance field at these regions is only regularized by inaccurate modeling of boundary condition and usually cause the wrong reconstruction. These problems cause significant larger errors compared with our methods by measuring the distance from the scan to the point cloud. Our method provides more accurate reconstruction and is robust to large holes in scans.

\section{Conclusion}
We present a robust, scalable, and accurate surface reconstruction algorithm that guarantees to provide watertight manifold for triangle soups. We specifically deal with challenges of orientation ambiguity and thin structures commonly existing in available 3D data. We use a volumetric representation to extract surfaces between exterior and occupied voxels and thereby remove orientation ambiguity.

In the future, there are several things that can be considered to further improve manifold remeshing. First, now we fix a hole if its size is smaller than a voxel size. However, the artist created CAD models could have big holes. It is interesting to explore semantic information in order to decide whether to fix the holes. Second, our method is not designed for noisy scanning data. It is a promising direction to combine our method jointly with denoising for better surface reconstruction.

\bibliographystyle{ACM-Reference-Format}
\bibliography{sample-base}

\clearpage
\section*{Appendix}
\begin{appendix}
\section{Build Octree}
\label{sec:appendix-octree}
Detailed algorithm for building initial octree from the reference mesh is shown in Algorithm~\ref{alg:construct-octree}, where the initial call sets $\mathcal{F}=\mathcal{F}_r$ and depth as $0$.
\begin{algorithm}[b]
\DontPrintSemicolon % Some LaTeX compilers require you to use \dontprintsemicolon instead
\KwIn{$\mathcal{F}$, Octree Node $T$, Current depth $d$, Target tree height $H$}
\KwOut{Constructed Octree $T$}
\If {$d=0$} {
    T.volume $\leftarrow [-1.1,1.1]^3$\\
    T.status $\leftarrow$ Occupied\\
    T.level $\leftarrow d$\\
    T.children $\leftarrow$ Node[8]\\
}
\If {$d<H$} {
    \For {$i \leftarrow 1$ to $8$} {
        C.volume $\leftarrow$ $i$-th half-cube of T.volume\\
        C.status $\leftarrow$ Empty\\
        C.level $\leftarrow d+1$\\
        $\mathcal{F}_C\leftarrow \{f | f\in\mathcal{F} \mathrm{\;and\;intersects\;C.volume}\}$\\
        \If {$|\mathcal{F}_C| > 0$} {
            C.status $\leftarrow$ Occupied\\
            ConstructVolume($\mathcal{F}_C$, C, $d+1$, $d^*$)
        }
        T.children[i] = C
    }
}
\Return{$T$}\;
\caption{ConstructVolume}
\label{alg:construct-octree}
\end{algorithm}

\section{Build Octree Connections}
\label{sec:appendix-connection}
We build internal connections in the octree recursively by calling ConnectOctree in Algorithm~\ref{alg:construct-internal-connection}, where it recursively builds internal connections in the children and additionally build external connections between neighboring children in Algorithm~\ref{alg:construct-external-connection}.
\begin{algorithm}[b]
\DontPrintSemicolon % Some LaTeX compilers require you to use \dontprintsemicolon instead
\KwIn{Octree Node $T$}
\KwOut{$T$ with Connections}
\If {$d<H$} {
    \For {$i \leftarrow 1$ to $8$} {
        ConnectOctree(T.children[i])
    }
    $P_x \leftarrow \{(1,2),(3,4),(5,6),(7,8)\}$\\
    $P_y \leftarrow \{(1,3),(2,4),(5,7),(6,8)\}$\\
    $P_z \leftarrow \{(1,5),(2,6),(3,7),(4,8)\}$\\
    \For {p$\in P_x$} {
        ConnectNodes(T.children[p.x], T.children[p.y],1)
    }
    \For {p$\in P_y$} {
        ConnectNodes(T.children[p.x], T.children[p.y],2)
    }
    \For {p$\in P_z$} {
        ConnectNodes(T.children[p.x], T.children[p.y],3)
    }
}
\Return{$T$}\;
\caption{ConnectOctree}
\label{alg:construct-internal-connection}
\end{algorithm}
\begin{algorithm}
\DontPrintSemicolon % Some LaTeX compilers require you to use \dontprintsemicolon instead
\KwIn{Neighboring Node $A$ and $B$, and connecting direction $d$}
\KwOut{$A$ and $B$ with connections}
\If {Both A and B are leaves} {
    Add an edge between A and B.
}
\Else {
    $I_A \leftarrow [[2,4,6,8],[3,4,7,8],[5,6,7,8]]$\\
    $I_B \leftarrow [[1,3,5,7],[1,2,5,6],[1,2,3,4]]$\\
    \If {A is leaf} {
        $N_A \leftarrow [A,A,A,A]$
    }
    \Else {
        $N_A \leftarrow$ A.children[$I_A$[d]]
    }
    \If {A is leaf} {
        $N_B \leftarrow [B,B,B,B]$
    }
    \Else {
        $N_B$ $\leftarrow$ B.children[$I_B$[d]]
    }
    \For {$i \leftarrow 1$ to 4} {
        ConnectNodes($N_A$[i], $N_B$[i])
    }
}
\Return{$T$}\;
\caption{ConnectNodes}
\label{alg:construct-external-connection}
\end{algorithm}

\section{More results on ModelNet10}
\begin{figure*}
\includegraphics[width=0.7\linewidth]{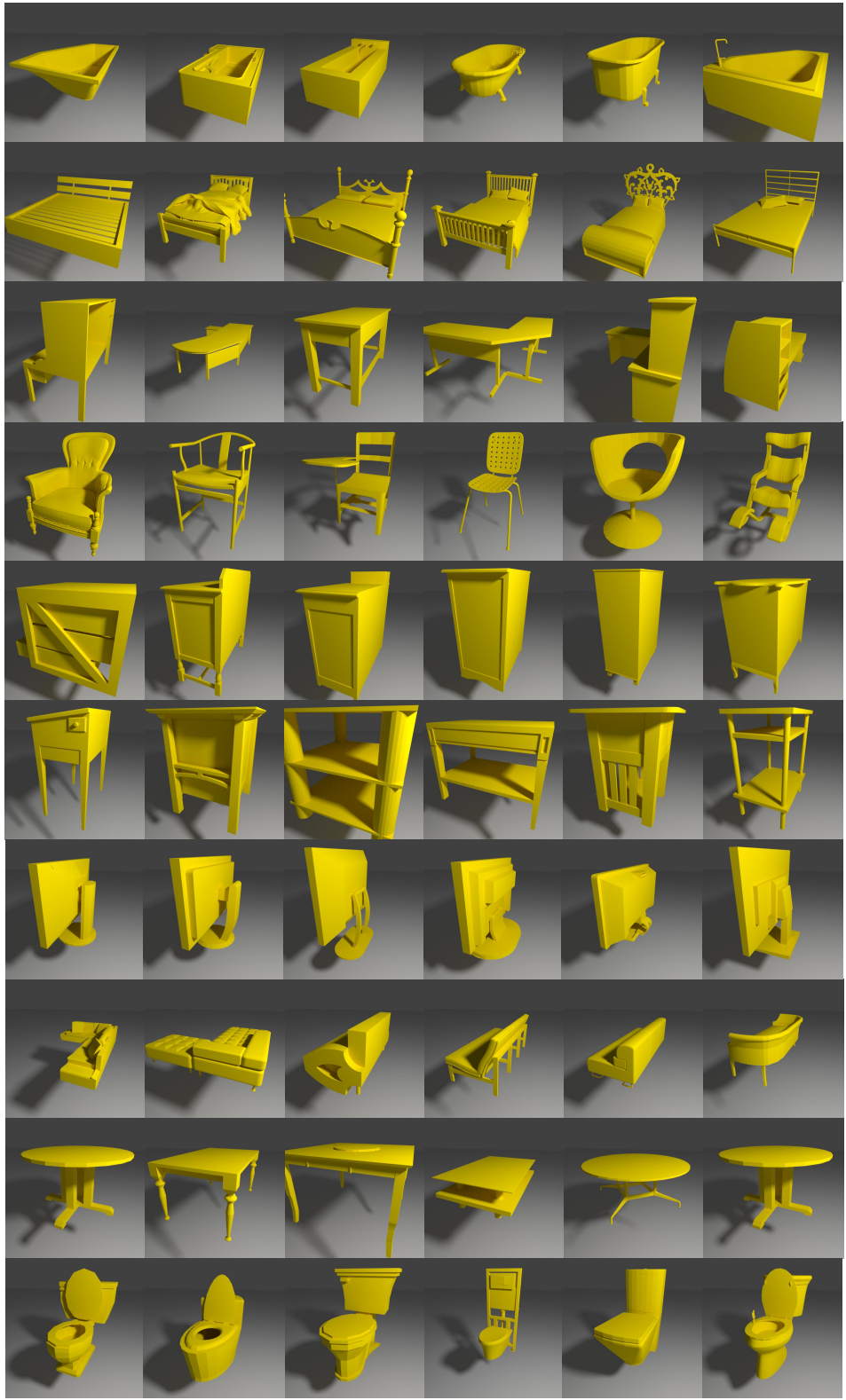}
\caption{Per-category ModelNet10 manifold results generated using our method.}
\label{fig:supp-modelnet}
\end{figure*}

We provide more manifold results generated by our method in the ModelNet10 dataset~\cite{wu20153d} in Figure~\ref{fig:supp-modelnet} for each category.
\end{appendix}

\end{document}